# Estimation of Rain Attenuation of Earth-to-Satellite Link over Nepal for $K_u$ & $K_a$ Bands

Jasmul Alam
*Qatar international Electrical co.*
Doha, Qatar
jasmulalam@gmail.com

Md. Sakir Hossain
*Dept. of Computer Science*
*American International University-Bangladesh*
Dhaka, Bangladesh
sakir.hossain@aiub.edu

Jauwad Ansari
*Almana Motors Company*
Doha, Qatar
Jauwad54321@gmail.com

Abu Zafar Md. Imran
*Dept. of Electronics and Telecommunication Engineering*
*International Islamic University Chittagong*
Chittagong, Dhaka
azmimran28@gmail.com

Imtiaz Kamrul
*Dept. of Electronics and Telecommunication Engineering*
*Internationall Islamic University Chittagong*
Chittagong, Dhaka
imtiazshuvo10@gmail.com

*Abstract*—Due to the extensive use of lower frequency bands & huge demand of large bandwidth in satellite communications, engineers need to use the higher frequency bands such as $K_u$ to $K_a$ bands. However, the rain attenuation is severe in these bands. Before installing an earth station, the estimation of the rain attenuation is a prerequisite task to know the required fade margin to ensure a certain availability of the network. In this paper, we estimate the rain attenuation for different regions of Nepal. The R-H and ITU models are used for rain rate and rain attenuation estimation, respectively. A significant temporal and spatial variation in rain attenuation is observed. Among three different regions of Nepal, namely Terai, Hilly, and Himalaya, while the maximum rain attenuation is found in Terai region, the minimum is in Himalaya. Jhapa of the Terai region experiences the highest attenuation and requires 80 dB and 24 dB fade margin for $K_a$ and $K_u$ bands, respectively. Solukhumbu of Himalaya region, on the other hand, is found to be the least rain attenuation affected site. The required fade margin for an earth station site in Solukhumbu for using $K_u$ and $K_a$ bands are 12 dB and 40 dB, respectively. The minimum attenuation, which is observed in November and December, is several times lower compared to the highest rain attenuation, observed in July. The minimum attenuation caused by the $K_a$ band is higher than the maximum attenuation caused by the $K_u$ band irrespective of the locations of the earth station sites.

## I. INTRODUCTION

World has seen an unprecedent growth in telecommunication networks in last two decades. Wireless communication has experienced the most development among the fields of the telecommunications. The reason behind the popularity of wireless communication is due to its capability to make communication services ubiquitous. A significant share of such systems is largely attained by dint of satellite communication, which has been witnessing an uptrend in its popularity. The satellite communication is particularly needed for long-distance communications and in rough terrain structure, hilly difficult-to-reach region in particular. Nepal is a landlocked small country, which is one of the heaviest mountainous country of the earth. Most of the regions are not accessible by the terrestrial communication system because of the mountains. About 75% of the land surface of Nepal is highly hilly area, where any kind of terrestrial communication is very difficult to install. However, these regions are popular tourist destinations, and the Nepalese economy largely depends on the tourist turnover. To facilitate the tourists, it is imperative to connect these regions to the telecommunication networks, and satellite communication is the only viable solution for providing such communication services to these regions.

In satellite communication, the C-band is widely used because of its low atmospheric attenuation. Popularity of satellite communication and the dramatic rise of multimedia services make the C-band nearly occupied. To support more earth stations and high-resolution video services, it is required to look for more bandwidth. In this regards, $K_u$ and $K_a$ bands are started to be used across the world for satellite communication. The usage of the higher frequency bands can help us get rid of the problem. However, such solution is not a complete blessing because the higher frequency bands are accompanied by an increased attenuation due to hydrometeors [1]. Some typical examples of the hydrometeors are raindrops, snowflakes, hail, ice-crystals and so on. Before using the $K_u$ and $K_a$ bands, it is necessary to first investigate the impact of the higher frequency on the quality of the earth-to-satellite (ETS) link. The impact depends on the climate of the place where the earth station is located. Since the climate varies significantly over the regions of the earth surface, the attenuation an ETS experiences also varies. In [2], the impacts of atmospheric scintillation in Bangladeshi climatic conditions are investigate. The rain attenuation because of the $K_u$ and $K_a$ bands for ETS links are investigated in [3], [4] for Bangladeshi climatic condition. In [5], the rain attenuation experienced by an ETE link for EHF (V and W) bands are predicted for Bangladeshi climate. Similar estimation of attenuation caused by hydrometeors are carried out in [6]-[9] for Indian, Malaysian, South African, and Ethiopian climates. The predictions find that the use of the $K_u$ and $K_a$ bands requires increased fade margin to maintain ETS link quality, and the required fade margin varies significantly from region to region. Since Nepal has considerably high diversity in climate because of its different terrain structure, estimating attenuation due to the hydrometeors is very significant. Since the attenuation caused by the rain is far higher among the hydrometeors, the most important is to estimate the rain attenuation.

In this paper, we estimate rain attenuation in Nepalese region for $K_u$ and $K_a$ bands. First, we estimate rain rate in three different regions of Nepal. Next, we estimate rain attenuation for different percentage time of a year. The rain attenuation variation in different months is also investigated.

## II. RAIN ATTENUATION PREDICTION MODEL

There are several rain attenuation prediction models [10]. The ITU model [11] and CCIR model [12] are applicable for world-wide climate, and the rest of the models in [10] are suitable only for specific regions of the earth. Since the ITU model is more comprehensive and accurate compared to the other models, we will use this model for estimating the rain attenuation in Nepal. The details of the model are noted below:

**Step 1:** Determine the rain height at the ground station of interest

$$h_R = h_0 + 0.36 \text{ (km)} \tag{1}$$

where $h_0$ is the freezing height above the sea level.

**Step 2:** Calculate the slant-path length

$$L_s(\theta) = \begin{cases} \frac{(h_R - h_S)}{\sin \theta}, & \theta \geq 5^0 \\ \frac{2(h_R - h_S)}{\left[\sin \theta^2 + \frac{2(h_R - h_S)}{R_e}\right]}, & \theta \leq 5^0 \end{cases} \tag{2}$$

where $h_R$ is the rain height (in km), from Step 1; $h_S$ is the altitude of the ground station site from the sea level (km), which is also known as elevation, El; $\theta$ is the elevation angle; and $R_E = 8500$ km (effective earth radius). $L_s$ can be negative when the rain height is smaller than the altitude of the ground station site. If a negative value occurs, $L_s$ is set to zero.

**Step 3:** The slant path length's horizontal projection, $L_G$, can be calculated as follow:
$$L_G = L_s \cos(\theta) \text{ km} \tag{3}$$

**Step 4:** Estimate the rain rate, $R_{0.01}$, for 0.01% time of a year with 1-minute integration time.

**Step 5:** Calculate the specific attenuation as

$$\gamma_R = K(R_{0.01})^\alpha \text{ dB/km} \tag{4}$$

**Step 6:** Next we calculate the horizontal reduction factor, $r_{0.01}$, which is determined from the rain rate $R_{0.01}$ as follows:

$$r_{0.01} = \frac{1}{1 + 0.78\sqrt{\frac{L_G \gamma_R}{f}} - 0.38(1 - e^{-2L_G})} \tag{5}$$

**Step 7:** Calculate the vertical adjustment factor, $V_{0.01}$ for 0.01% of the time as

$$V_{0.01} = \frac{1}{1 + \sqrt{\sin \theta}\left[31\left(1 - e^{-\left(\frac{\theta}{1+X}\right)}\right)\frac{\sqrt{L_G \gamma_R}}{f^2} - 0.45\right]} \tag{6}$$

$$L_R = \begin{cases} \frac{L_G r_{0.01}}{\cos \theta}, & \zeta > \theta \\ \frac{(h_R - h_S)}{\sin \theta}, & \zeta \leq \theta \end{cases} \tag{7}$$

and

$$\zeta = \tan^{-1}\frac{(h_R - h_S)}{(L_G r_{0.01})}(deg) \tag{8}$$

$$X = \begin{cases} 36 - |\varphi| & \varphi < 36 \\ 0 & \varphi \geq 36 \end{cases} \tag{9}$$

**Step 8:** Next, the effective path length, $L_E$, is determined as

$$L_E = L_R V_{0.01} \text{ (km)} \tag{10}$$

**Step 9:** Calculate the attenuation exceed for 0.01% of an average year. The predicted attenuation exceeded for 0.01% of an average year, $A_{0.01}$, is determined from

$$A_{0.01} = \gamma_R L_E \text{ (dB)} \tag{11}$$

The value of $A_{0.01}$ is only for 0.01% for an average year. To calculate the attenuation exceeded for other percentage of time, the following equation needs to be used:

$$A_p = A_{0.01}\left(\frac{p}{0.01}\right)^{-[0.655 + 0.033 \ln(p) - 0.045 \ln(A_{0.01}) - \beta(1-p)\sin\theta]} \tag{12}$$

where

$$\beta = \begin{cases} 0, & for\ p \geq 1\%\ or\ |\varphi| \geq 36^o \\ -0.005(|\varphi| - 36), & for\ p < 1\%\ or\ |\varphi| < 36^o \\ -0.005(|\varphi| - 36) + 1.8 - 4.25\sin\theta, & Otherwise \end{cases}$$

This method provides an estimate of long-term statistics due to rain. Large year-to-year variability in rainfall statistics can be expected when the predicted results are compared with the measured statistics.

## III. RAINFALL RATE ESTIMATION

A great variation in climate is observed in Nepal because of the extremely large range of altitude from north-to-south distance. According to [13], there are six climatic regions in Nepal depending on the altitudes as is given in Table I.

While the Himalayan region experiences the Arctic climate, the Terai region experiences the warm tropical climate. The rest of the climates exist in the hilly region. The rainfall is maximum in the eastern Nepal, and decreases as we proceed to the west.

TABLE I: CLIMATIC REGIONS IN NEPAL

| Climatic regions | Altitude (in meters) |
|---|---|
| Arctic | Over 5,500 |
| Sub-Arctic | 4,500 to 5,500 |
| Cold Temperature | 3,500 to 4,500 |
| Cool Temperature | 2,000 to 3,500 |
| Warm Sub Tropical | 700 to 2,000 |
| Warm Tropical | Below 700 |

While a maximum of 2,500 millimeters (mm) of rain is received in the eastern Nepal, the minimum of 1,000 mm is received in the western zone. The central region, Kathmandu, receives 1,420 mm rain annually. Since the rainfall does not occur equally around the year, the rain rate varies significantly over time and space. For this reason, we use cumulative distribution (CD) of rain to find the rate of rain for a particular percentage of time of an average year. The annual and monthly rainfall data in different cities of Nepal are collected from the Meteorological Forecasting Division of Nepal. To compute the rainfall rate, we use Rice-Holmberg rain model (R-H rain model) [14]. Parameters of the model include average annual rainfall depth, M and the thunderstorm ration, , which is the ratio between the convective rain amount and annual rainfall. According to the map in [14], $\beta = 0.7$ for Nepal. The parameters of the R-H model and ITU rain attenuation models are TABLE II.

TABLE II: PARAMETERS FOR R-H RAIN MODEL AND ITU RAIN ATTENUATION MODEL

| Region | City | Latitude/ Longitude | El (meters) | M |
|---|---|---|---|---|
| Terai | Janakpur | 26°71'N 85°92'E | 75.76 | 1431 |
| | Biratnagar | 26°48'N 87°28'E | 81 | 1550 |
| | Birgunj | 27°01'N 84°87'E | 85.29 | 1611 |
| | Jhapa | 26°55'N 88°08'E | 93.24 | 2112 |
| Hilly | Kathmandu | 27°70'N 85°31'E | 1301.93 | 1841 |
| | Pokhara | 28°26'N 83°97'E | 989.69 | 2030 |
| | Dhankutta | 26°95'N 87°33'E | 682.09 | 1809 |
| | Okhaldhunga | 27°19'N 86°30'E | 1561 | 1460 |
| Himalayan | Namche | 27°49'N 86°43'E | 3440 | 3440 |
| | Mustang | 28°47'N 83°43'E | 2743 | 2743 |
| | Solukhumbu | 27°30'N 86°35'E | 3700 | 3700 |
| | Humla | 29°77'N 82°19'E | 2373 | 2373 |

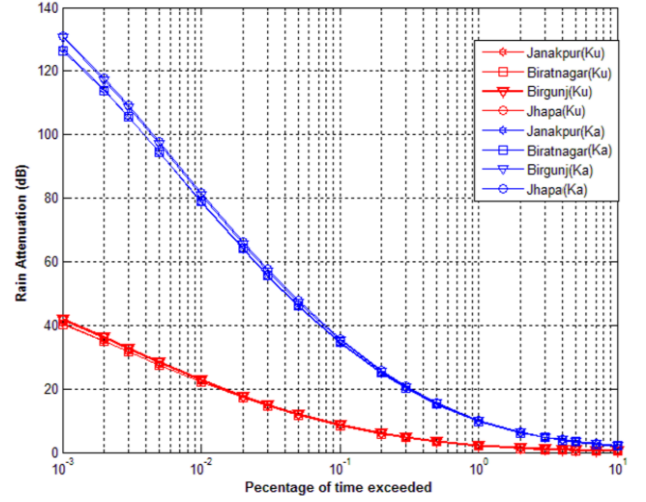

Fig. 1 Rain attenuation in Terai region for different percentage of time of the year.

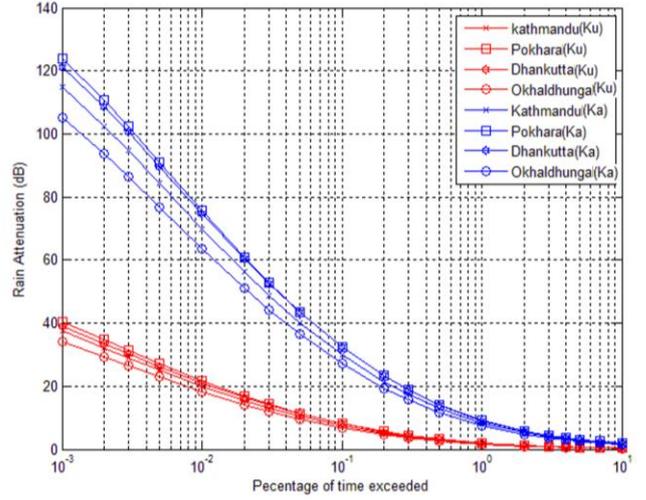

Fig. 2 Rain attenuation in Hilly region for different percentage of time in a year

## IV. RAIN ATTENUATION ESTIMATION

We will investigate the rain attenuation in the three regions of Nepal. In each region, the four cities as are mentioned in Table II will be considered. Furthermore, the monthly variation of the attenuation in the cities will also be investigated. First, the rain attenuation in Terai region is investigated in Fig. 1. The attenuation is almost similar in all the cities of the region, with a difference of about 5 dB for $K_a$-band. A huge difference is observed as the frequency band changes from $K_u$ to $K_a$. At 0.01% time of the year, the attenuation is about four times higher in $K_a$-band compared to that in the $K_u$-band. While the attenuation is 23 dB in the $K_u$-band, it goes as high as 80 dB in the $K_a$-band. The difference becomes 80 dB at 0.001% time of a year.

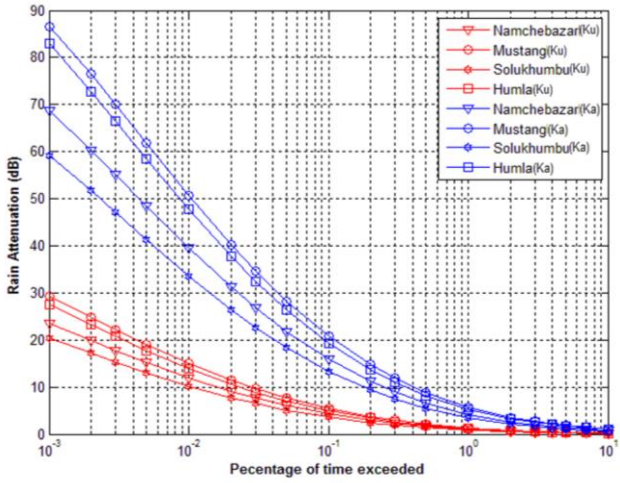

Fig. 3 Rain attenuation in Himalaya region for different percentage of time in a year

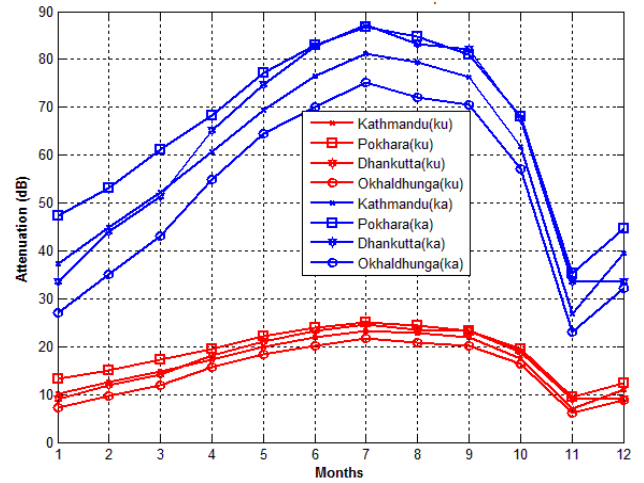

Fig. 5 Monthly rain attenuation in $K_a$ and $K_u$ bands in Hilly region.

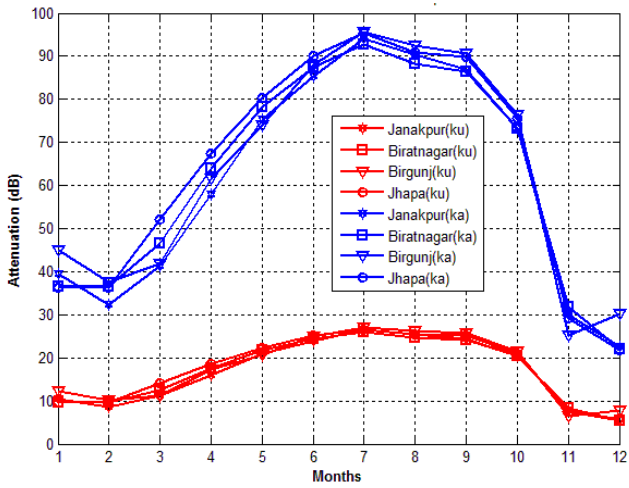

Fig. 4 Terai region's rain attenuation in different months of a year

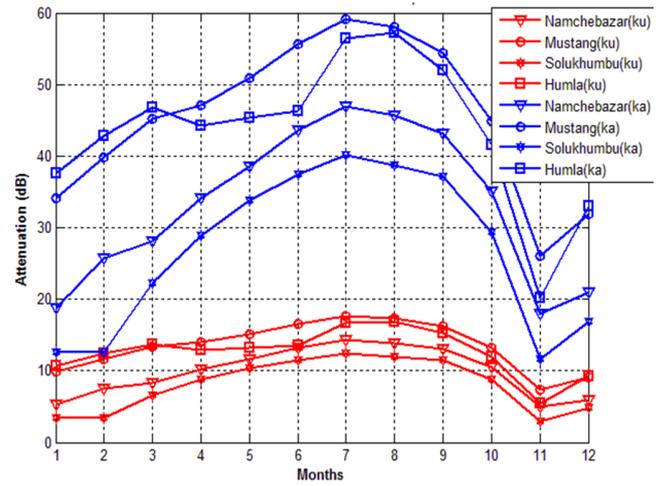

Fig. 6 Attenuation in $K_a$ and $K_u$ bands in Himalayan region.

Figure 2 shows the rain attenuation for the Hilly region. As is seen, there is a significant variation in attenuation among the cities of the region. Pokhara suffers most severely compared to the other cities. On the other hand, Okhaldhunga experiences the least attenuation. A satellite signal in Pokhara losses the signal strength of 15 dB compared to Okhaldhunga where the attenuation is 23 dB. All cities experience almost equal attenuation of around 20 dB in $K_u$-band. The rain attenuation for the cities of Himalaya region is shown in Fig. 3. Similar to the Hilly region, the difference in attenuation for $K_a$-band experienced in this region is about 15 dB for 0.01% percent of a year, with a maximum of 50 dB is found in Mustang and a minimum of 35 dB is observed in Solukhumbu. For the $K_u$-band, it is 5 dB. Comparing Fig. 1, 2, and 3, we see that Terai region suffers from the maximum attenuation, while the Himalaya region is the least affected region. The maximum attenuations are 80 dB, 75 dB and 50 dB in the Terai, Hilly and Himalaya regions, respectively. In Nepal Jhapa and Birganj experience the highest attenuation, and the signal attenuation is the least in Solukhumbu. For the same band, a satellite signal in Jhapa or Birganj attenuated twice the amount compared to that of Solukhumbu.

Rainfall does not occur equally throughout a year; rather it varies significantly from season to season. The monthly variation of attenuation due to rain will be investigated next. Fig. 4 shows monthly variation of the rain attenuation in Terai region. There is no considerable difference in monthly attenuation among the cities in Terai region. There is a huge difference between the attenuation in $K_u$ and $K_a$ bands; in rainy season, the attenuation in $K_a$-band is almost three times higher compared to that in $K_u$-band. In $K_a$-band, the minimum attenuation of about 22 dB is observed in December, then it slightly increases in next two months. From March, it steadily increases until reaches 95 dB in July after which it starts falling sharply and reaches to the minimum in December. In the $K_u$-band, the minimum and maximum attenuations are 17 dB and 27 dB respectively.

The cities in the Hilly region experience varying degree of rain attenuation for a particular month as is shown in Fig. 5. Unlike the Terai region, the minimum attenuation of 22 dB (in Okhaldhunga) to 35 dB( in Pokhara and Dhankutta) and 17 dB (Okhaldhunga) to 20 dB (in Pokhara) are found in November in $K_a$ and $K_u$ bands, respectively. The maximum attenuation varies 20 dB and 5 dB among the cities in the $K_a$ and $K_u$ bands, respectively. About 50 dB variation in attenuation is observed throughout a year for $K_a$-band, while it is 17 dB for $K_u$-band. Fig. 6 shows the monthly rain attenuation variation for different cities of Himalaya region of Nepal. For $K_a$-band, the maximum attenuation found in July varies 20 dB from 40 dB in Solukhubu to 60 dB in Mustang, while the minimum figures (found in November) of the cities are 12 dB and 26 dB respectively. The variation among the cities throughout a year for the $K_u$-band is 5 dB. Figs. 4, 5 and 6 are obtained using vertical polarized antennas. Through extensive simulation, it is found that there is about 3 dB reduced attenuation for horizontal polarization

## V. CONCLUSION

In this paper, the rain attenuation experienced by an ETS link is estimated for different cities of Nepal. A significant temporal and spatial variation of the attenuation is observed. Among the three regions of Nepal, while the Terai region experiences the highest attenuation, the Himalaay region is the least affected region. To ensure 99.99% of network availability with $K_a$-band, the amount of fade margin should be at least 80 dB, 62 dB and 34 dB in Terai, Hilly and Himalaya regions, respectively. For $K_u$-band, the figures are 32 dB, 18 dB and 10 dB, respectively. The most favorable cities in terms of fade margin for installing an earth station in Terai, Hilly and Himalaya regions are Biratnagar, Okhaldhunga and Solukhumbu, respectively. On the other hand, Jhapa, Pokhara and Mustang are the least favorable cities, respectively. Among all the cities of Nepal, Solukhumbu requires the least fade margin, and Jhapa needs the highest fade margin, on the other hand.

The attenuation due to rain can be kept to the minimum level by using horizontal polarization, or using $K_u$-band. In addition, we can also install an earth station in the city of Solukhumbu of Himalaya region, as the rain attenuation is found minimum there. Then, a backhaul network can be used to send the received signal to other regions from where the signal can be retransmitted over terrestrial networks. In this way, the overall rain attenuation can be minimized in the downlink. Nevertheless, this network architecture will face another problem from the ice attenuation, though such attenuation is not significant. But due to heavy snowfall in Himalaya region, hitting mechanism must be introduced at the parabolic reflector antennas.

In this paper, the rain attenuation is predicted for different regions considering different parameters. However, there are still scopes to investigate other issues. Firstly, the rain depolarization can be estimated. Secondly, the tropospheric scintillation due to the change of temperature and relative humidity can be computed. Finally, the site diversity scheme can be investigated to find the gain due to change of earth station site.

REFERENCES

[1] M.S. Hossain, "Rain attenuation prediction for terrestrial microwave link in Bangladesh", Journal of Electrical and Electronic Engineering, vol. 7, no. 1, pp. 63-68, 2014.

[2] M.S. Hossain and M. A. Samad, "The tropospheric scintillation prediction of earth-to-satellite link for Bangladeshi climatic condition", Serbian Journal of Electrical Engineering, vol. 12, no. 3, pp. 263-273, 2015.

[3] A.Z.M. Imran, M.T. Islam, A. Gafur and Y. W. Rabby, "Rain attenuation prediction analysis and contour map design over Bangladesh", In Proc. International Conference on Computer and Information Technology, 2016.

[4] M. Kamruzzaman and M.S. Islam, "Rain attenuation prediction for satellite communications link at Ku and Ka bands over Bangladesh", In Proc. International Conference on Electrical Engineering and Information Communication Technology (ICEEICT), 2014.

[5] M. S. Hossain and M. A. Islam, "Estimation of rain attenuation at EHF bands for earth-to-satellite links in Bangladesh", In Proc. International Conference on Electrical, Computer and Communication Engineering, pp. 589-593, 2017.

[6] S. K. Sarkar, R. Kumar, and J. Das, "Rain Attenuation at 13 GHz over a LOS Terrestrial Link Situated in Indian Eastern Sector", Indian Journal of Physics, vol. 77B, no. 3, pp. 271-275, 2003.

[7] U. Kesavan, A. R. Tharek, A. Y. Abdul Rahman, and S.K. Abdul Rahim, " Comparative Studies of the Rain Attenuation Predictions for Tropical Regions", Progress In Electromagnetics Research M, vol. 18, 17-30, 2011.

[8] P. A. Owolawi, S. J. Malinga, and T. J. O. Afullo, "Estimation of Terrestrial Rain Attenuation at Microwave and Millimeter Wave Signals in South Africa Using the ITU-R Model, ",PIERS Proceedings, pp. 27-30, 2012.

[9] F. D. Diba, T. J. Afullo, "Estimation of rain attenuation over microwave and millimeter bands for terrestrial radio links in Ethiopia", AFRICON, 2015.

[10] G. S. Feldhake and L. A. Sengers, "Comparison of multiple rain attenuation models with three years of Ka band propagation data concurrently taken at eight different locations", Journal of Space Communication, vol. 2, Fall 2002.

[11] Recommendation ITU-R P.618-10 (10/2009), "Propagation data and prediction methods required for the design of earth-space telecommunication System", P Series.

[12] CCIR Report 564-3, 1986.

[13] S. S. Negi, Forest and Forestry in Nepal, South Asia Book, p.12, 1984.

[14] P. L. Rice and N. R. Holmberg, "Cumulative time statistics of surface point rainfall rates", IEEE Trans. on Communications, vol. 21, no. 10, pp. 1131-1136, 1973.